\documentclass[runningheads]{llncs}

\usepackage[hidelinks]{hyperref}
\usepackage{graphicx}
\usepackage{amsmath}
\usepackage{pgfplots}
\pgfplotsset{compat=1.16} 

\usepackage{booktabs}

\usepackage{siunitx}

\usepackage{xcolor} 
\definecolor{azure}{rgb}{0.0, 0.5, 1.0}
\definecolor{darkgreen}{rgb}{0.0, 0.5, 0.0}
\definecolor{amaranth}{rgb}{0.9, 0.17, 0.31}
\definecolor{cadetgrey}{rgb}{0.57, 0.64, 0.69}
\definecolor{aureolin}{rgb}{0.99, 0.93, 0.0}

\usepackage{listings}

\usepackage[caption=false]{subfig}

\usepackage{tikz}
\usetikzlibrary{shapes}

\usepackage{todonotes}
\begin{document}
\title{HPX with Spack and Singularity Containers: Evaluating Overheads for HPX/Kokkos using an astrophysics application
}

\titlerunning{HPX with Spack and Singularity Containers}

%
%
\author{
Patrick Diehl\inst{1,2}\orcidID{0000-0003-3922-8419} \and Steven R. Brandt\inst{1} \and Gregor Dai\ss{}\inst{3} \and Hartmut Kaiser\inst{1}\orcidID{0000-0002-8712-2806}}
\authorrunning{P. Diehl et al.}
%
\institute{Center of Computation \& Technology, Louisiana State University
\email{\{pdiehl,hkaiser,sbrandt\}@cct.lsu.edu}\\
\and Department of Physics and Astronomy, Louisiana State University  \and Institute for Parallel and Distributed Systems, University of Stuttgart \email{Gregor.Daiss@ipvs.uni-stuttgart.de} }
\maketitle              
\begin{abstract}
Cloud computing for high performance computing resources is an emerging topic. This service is of interest to researchers who care about reproducible computing, for software packages with complex installations, and for companies or researchers who need the compute resources only occasionally or do not want to run and maintain a supercomputer on their own. The connection between HPC and containers is exemplified by the fact that Microsoft Azure's Eagle cloud service machine is number three on the November 23 Top\ 500 list. For cloud services, the HPC application and dependencies are installed in containers, \emph{e.g.}\ Docker, Singularity, or something else, and these containers are executed on the physical hardware. Although containerization leverages the existing Linux kernel and should not impose overheads on the computation, there is the possibility that machine-specific optimizations might be lost, particularly machine-specific installs of commonly used packages. In this paper, we will use an astrophysics application using HPX-Kokkos and measure overheads on homogeneous resources, e.g.\ Supercomputer\ Fugaku, using CPUs only and on heterogenous resources, \emph{e.g.}\ LSU's hybrid CPU and GPU system. We will report on challenges in compiling, running, and using the containers as well as performance performance differences.

\keywords{C\texttt{++}  \and HPX \and AMT \and Parallelism.}
\end{abstract}

\section{Introduction}
In recent years, cloud computing for high performance computing resources gained more interest. The most recent traction was that Microsoft's Aszure Eagle cloud service machine is number three on November 23 Top\ 500 list. Another example was The Salishan 2023 conference where the gains and losses of HPC applications in the cloud and opportunities and challenges in adopting cloud software technologies for HPC were discussed. In this paper, we use cloud software technologies, like Singularity, to compile and run our astrophysics HPC application. We look into the potential gains of using containers for HPC applications. On the other side, we look into potential losses for using containerization. One common question for containerization is the performance difference compared to running on the host.  

Before we can measure any performance differences, we must compile our HPC application. The original workflow was to SSH to the head node and use either module files to load compilers and dependencies and use a standard build system, \emph{e.g.}\ Cmake or Make. Another option is the usage of Spack or EasyBuild as HPC packet managers (we do not wish to give the impression that the installation of a complex software application is ever easy). However, when using containers, our workflow is different. Our preferred option was to generate a Docker file and compiler Octo-Tiger using Spack. Since Docker requires root permission this step was outsourced to our local machines. The compiled image was converted using singularity build on the cluster.

One attractive property of containerization is that it offers better reproducibility~\cite{9882991}. To reproduce the results in a published paper many details, \emph{e.g.}\ compiler version, software versions, and input files, must be carefully documented. The Supercomputing (SC) conference series introduced the reproducibility initiative, requiring detailed provenance and documentation on compiling, installing, and running the software~\cite{9547695}. However, sometimes compiler versions and libraries change, for example, due to updates of the supercomputer. If the container is archived, it can be later used to run the HPC application with the same compiler and library versions. Running on the same hardware is a different story if the supercomputer is decommissioned.

The paper is structured as follows: Section~\ref{sec:related:work} discusses the related work. Section~\ref{sec:software:stack} introduces the software stack. Section~\ref{sec:workflow} emphasizes the workflow to compile and run HPX applications within containers. Section~\ref{sec:performance:differences} investigates the performance differences. Finally, Section~\ref{sec:conclusion} concludes the paper.

\section{Related Work}
\label{sec:related:work}
There have been many studies of the impact of containers on HPC codes.
The performance of Docker containers was studied in~\cite{casalicchio2017measuring,rad2017introduction,10.1145/3219104.3229280,sparks2019enabling}. The usage of Docker in HPC applications was studied~\cite{7562612,alles2018assessing,chung2016using,10.1007/978-3-319-20119-1_36}. A comparison of virtualization and containers was done in~\cite{7921010}. Sarus, a container engine for HPC environments was presented in~\cite{benedicic2019sarus}.
The integration of MPI with Docker containers is described in~\cite{7923810,7923813}. Production runs using containers in biological simulations were conducted in~\cite{8820966}. A representative study of
state-of-the-art container solutions (Docker, Podman, Singularity, and Charliecloud) in HPC environments was presented in~\cite{9284294}. Overheads of computation and communication on Linux containers were documented in~\cite{8748885}. However, none of these investigated the overheads for asynchronous many-task runtime systems, like the C\texttt{++} standard library for parallelism and concurrency (HPX). Other distributed many-task runtime systems that could be studied are Charm\texttt{++}~\cite{kale1993charm++}, Unitah~\cite{germain2000uintah}, Chapel~\cite{chamberlain2007parallel}, PaRSEC~\cite{bosilca2013parsec}, and Legion~\cite{bauer2012legion}. For a detailed comparison of these systems (independent of any containerization considerations), we refer to~\cite{thoman2018taxonomy}.


\section{Software Stack}
\label{sec:software:stack}
In this section, we give a brief overview of Octo-Tiger, its most important dependencies and the dependency management.
\subsection{Notable Octo-Tiger Dependencies}

\subsubsection{HPX:}
HPX is an Asynchronous Many-Task Runtime System (AMT)~\cite{kaiser2020hpx}.
With it, we can manage data and execution dependencies within Octo-Tiger with a task graph built by using C\texttt{++} futures and continuations (\texttt{hpx::future}).
HPX can easily handle millions of tasks, which are being processed by just a few (usually one per CPU core) HPX worker threads.
Beyond futures, HPX implements the C\texttt{++}20 API for parallelism and concurrency, for example including functionality such as \texttt{hpx::mutex}.
HPX also comes with distributed capabilities: We can asynchronously call methods on HPX components residing on other compute nodes, getting futures in return to integrate these calls into the task graph.
HPX supports the same syntax and semantics for these remote function calls as for local function calls by using an Active Global Address Space (AGAS) underneath.
HPX further offers multiple communication backends, based on either TCP, MPI or LCI.
Furthermore, HPX offers tight integrations with CUDA, ROCm and, recently, also SYCL~\cite{10.1145/3585341.3585354}, allowing us to integrate asynchronous GPU kernels and CPU/GPU data-transfers into the task graph as well.

\subsubsection{Kokkos and HPX-Kokkos:}
Kokkos is a framework for developing performance-portable compute kernels~\cite{9485033}.
With it, we can write a compute kernel once and run it on different execution and memory spaces, depending on the target device.
These spaces are available for all major platforms. For example, there are CUDA execution and memory spaces available to target NVIDIA GPUs.
Notably, Kokkos also contains an HPX execution space, which allows Kokkos kernels to run on HPX worker threads, eliminating the need for conflicting thread pools (as we would encounter if we tried to use the OpenMP Kokkos execution space in an HPX application).
Kokkos also includes SIMD types to allow for explicit SIMD vectorization. These use the appropriate SIMD instructions when instantiated on the CPU (for example AVX512), while keeping the kernel compatible with GPUs by using scalar operations there~\cite{sahasrabudhe2019portable}.

HPX-Kokkos is an additional, thin compatibility layer between HPX and Kokkos~\cite{daiss2021beyond}.
Unlike the aforementioned HPX Execution Space within Kokkos (meant for running Kokkos kernels on the HPX worker threads), HPX-Kokkos allows us to treat Kokkos kernels themselves as HPX tasks.
This enables us to seamlessly integrate Kokkos kernels into the HPX task graph.
This is extremely useful for scheduling asynchronous continuations for Kokkos kernels, post-processing or communicating their results automatically once the respective kernel is done.

By now, all major compute kernels in Octo-Tiger have been ported to Kokkos and support using explicit SIMD vectorization with the aforementioned SIMD types if necessary.

\subsubsection{CPPuddle:}
CPPuddle is a utility library for task-based GPU programming, suited for use with HPX.
It provides special allocators for GPU buffers (reusing previous allocations wherever possible) and executors for work aggregation (kernel fusion) which can fuse together similar GPU kernels on-the-fly~\cite{daiss2022aggregation}.
Together, they help to avoid GPU device starvation when dealing with a multitude of small GPU kernels, making CPPuddle especially suited for adaptive, tree-based codes like Octo-Tiger.

\subsubsection{Other dependencies:}
In addition to the aforementioned dependencies, Octo-Tiger requires a number of other frameworks installed. We need hdf5 and silo to handle IO. We further use hwloc and jemalloc for efficiency, and we need Boost for its various utilities. Depending on the machine we target, we also need CUDA, ROCm, or OneAPI (for SYCL) installed.
For each of the aforementioned dependencies we also need to handle their respective dependencies.
Overall this set of dependencies often leads to us using slightly different dependency versions depending on what machine we run on as we have to work with the given modules and versions. 
\subsection{Octo-Tiger}
Oct-Tiger is an astrophysics application for the simulation of stellar mergers, written in C\texttt{++}~\cite{marcello2021octo}.
It simulates binary star systems where a mass transfer between the two stars occurs. Depending on the conditions in these systems, this mass exchange can lead to a merger which in turn can yield various interesting outcomes, such as a Type Ia supernovae. 
Previously, Octo-Tiger has been used to study R Coronae Borealis
stars and the merger of bipolytropic stars~\cite{kadam2018numerical}.
Earlier runs with Octo-Tiger were conducted on Piz Daint~\cite{10.1145/3295500.3356221} and Cori. Currently we are testing the code on Perlmutter and Fugaku~\cite{diehl2023simulating}.
Octo-Tiger's need for computational resources also makes cloud computing, and by extension containers, interesting for us.

Octo-Tiger models stars as self-gravitating, inviscid fluids.
Thus, we need two coupled solvers: We employ the Fast-Multipole-Method to get the gravitational field generated by the fluid, and we use finite volumes for the hydrodynamics solver.
Octo-Tiger uses an adaptive octree as its data-structure, with the resolution focusing on the atmosphere between the two stars where the mass transfer is happening.
For efficiency, each tree-node in Octo-Tiger contains an entire $8x8x8$ sub-grid (though this is configurable at compile time).

Octo-Tiger is completely built upon HPX, with each tree-node being an HPX component. 
HPX was an ideal choice here, as it makes distributed tree-traversals convenient from a developer's standpoint: For instance, we do not need to remember whether a tree child-node is located on the same compute node or not. We can invoke its functions all the same, with HPX taking care of communicating the call to the correct compute node, giving us a future in return. This allows for quickly and asynchronously building a task-graph for each tree-traversal, which crucially helps us to avoid resource starvation by making parallel work available as soon as possible in large, distributed runs.

Together with the performance-portability of the Kokkos kernels, these distributed HPX features allow Octo-Tiger to target both current GPU and CPU supercomputers efficiently.
\subsection{Build and Dependendency Management}
\subsubsection{Legacy Buildscripts:}
To manage builds with Octo-Tiger and all its dependencies, we previously used a set of custom bash build scripts~\footnote{\url{https://github.com/STEllAR-GROUP/OctoTigerBuildChain}}.
However, these build scripts became unwieldy over time as we targeted more machines and platforms, especially since this often meant adding more dependencies (for example, CUDA, ROCm, Kokkos, SYCL or explicit SIMD SVE types for Fugaku). 
This made the deployment on new Supercomputers possible but bothersome, as we usually had to manually adjust the build scripts yet again for each new machine while trying not to break any of the existing support for other platforms.

Ensuring the reproducibility of builds (and thus of the performance results) was challenging as well, as this involved manually fixing the build scripts to a certain commit, as well as outlining the exact way of invoking the scripts and listing all modules that have to be loaded. 
An example of this can be seen in the reproducibility appendix of~\cite{daiss2022simd}.

This ultimately prompted us to look for alternatives, leading us to Spack and, later on, Singularity.
\subsubsection{Spack:}
Spack~\cite{gamblin2015spack} is a package manager to build and install multiple versions and configurations of software. 
Spack allows the installation of packages in the user space and is, therefore, widely adopted in the HPC community. Another solution worth mentioning here is \textit{EasyBuild}. However, we decided on Spack due to the flexibility it provides.

Spack expresses a package's variants, dependencies, utilized compiler and target platforms all in one single string called Spack spec. 
When installing a package, the user provides a spec that allows them to extensively modify the package without having to change the package recipe itself. 
The Spack concretizer will turn this input spec into a complete, concretized spec (meaning it will discover a compatible set of dependencies that works with the given input spec according to the constraints within all involved packages). 
This concretized spec will then be built and installed together with its dependencies. 

This makes Spack a powerful tool to adjust packages to each machine in question as it allows us to work around problems by, for example, easily switching dependency versions, variants, and compilers or by disabling certain features.
For example, the input Spack spec used on Fugaku was:
\begin{lstlisting}[language=bash,caption=Spack command to compile Octo-Tiger on Supercomputer\ Fugaku]
spack -v  install -j 4 octotiger@0.10.0 +kokkos +kokkos_hpx_kernels simd_extension=SVE simd_library=STD build_type=Release %gcc@12.2.0 ^hpx malloc=jemalloc networking=none instrumentation=apex +generic_coroutines ^bzip2@1.0.6 ^git@2.39.1 ^silo~mpi
\end{lstlisting}
For brevity, we omit the concretized spec as it contains all dependencies, not just the ones we manually adjusted, and is thus extremely verbose.

Reproducibility is also streamlined with Spack (compared to our custom build scripts), too. We simply need to store the string with the concretized Spack spec for each machine, as well as the Spack version used. Even packages already provided on the system (by modules) will show up in this concretized Spack spec if they have been added as an external package to Spack previously.

One of the contributions of this paper is providing a Spack package for Octo-Tiger: Creating this Spack package for Octo-Tiger was eased by the fact that HPX, Kokkos, and HPX-Kokkos are already available as Spack packages. While we had to adapt those package recipes slightly (mainly to support our SYCL variant), the main chunk of work for this was to create the new CPPuddle and Octo-Tiger Spack package recipes, as they needed to support a multitude of versions and variants. 

The new Octo-Tiger package and all our modifications to other Spack packages are available in our Spack repository on GitHub \footnote{\url{https://github.com/G-071/octotiger-spack}}. 
The Spack package is already in use by our developers both on our local development machines, our university servers and Supercomputers such as Perlmutter and Fugaku.
We further integrated the Octo-Tiger Spack package in our CI pipeline as a Jenkins matrix job over a list of tuples, with each tuple consisting of a SLURM command and an associated Spack spec. This provides an easy way of testing all our relevant variants on different machines in a single Jenkins Pipeline.
\subsubsection{Singularity:}
Docker images~\cite{merkel2014docker} are widely used as containers. However, most docker commands require root access to be executed. Root access is possible on local development machines but not on HPC resources, like supercomputers or the cloud. Singularity can convert Docker images, however, and already existing ones can still be used. Unlike Docker, Singularity does not require root access which is essential for supercomputers or the cloud.

Spack provides experimental support for generating container files from its package recipes. These can be used with Docker and Singularity, which is something we were interested in trying out and will subsequently test in this work as well.

\section{Workflow}
\label{sec:workflow}

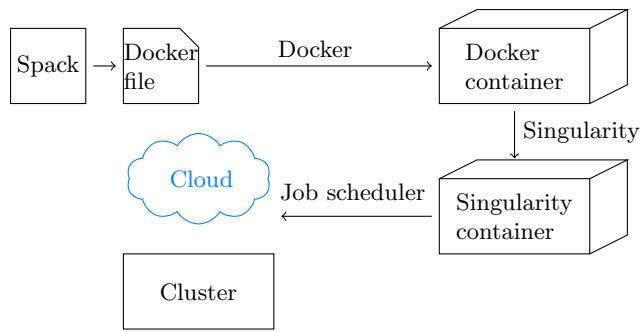
\begin{figure}[tb]
    \centering
    \begin{tikzpicture}
    \draw (-1.5,1) -- (-0.5,1) -- (-0.5,0) -- (-1.5,0) --cycle ;
    \node[align=left] at (-1,0.5) {Spack};
    \draw[->] (-0.4,0.5) -- (-0.1,0.5);
    \draw (0,0) -- (1,0) -- (1,0.75) -- (0.75,1) -- (0,1) -- cycle;
    \node[align=left] at (0.5,0.5) {Docker\\ file};
    \draw[->] (1.1,0.5) -- (4.1,0.5);
    \node[above] at (2.55,0.5) {Docker};
    \draw (4.2,0) -- (6.2,0) -- (6.2,1) -- (4.2,1) -- cycle;
    \draw (4.2,1) -- (4.7,1.25) -- (6.7,1.25) -- (6.2,1);
    \draw (6.7,1.25) -- (6.7,0.25) -- (6.2,0);
    \node[align=left] at (5.2,0.5) {Docker\\ container};
    \draw (4.2,-1) -- (6.2,-1) -- (6.2,-2) -- (4.2,-2) -- cycle;
    \draw (4.2,-1) -- (4.7,-.75) -- (6.7,-.75) -- (6.2,-1);
    \draw (6.7,-.75) -- (6.7,-1.75) -- (6.2,-2);
    \node[align=left] at (5.2,-1.5) {Singularity\\ container};
    \draw[->] (5.2,-0.1) -- (5.2,-0.73) ;
    \node[right] at (5.2,-0.375) {Singularity};
    \node [cloud, draw,cloud puffs=10,cloud puff arc=120, aspect=2, inner ysep=1em,azure] at (1,-1.) {};
    \node[right,azure] at (0.5,-1) {Cloud};
    \draw (0,-2.) -- (2,-2.) -- (2,-3) -- (0,-3) -- cycle;
    \node at (1,-2.5) {Cluster};
    \draw[->] (4.1,-1.5) -- (2.1,-1.5);
    \node[above] at (3.05,-1.4) {Job scheduler};
    \end{tikzpicture}
    \caption{Workflow to run in containers: A Docker file contains the information of the operating system and Spack instructions to compile Octo-Tiger. We use Docker to generate a Docker container using the Docker file. The Docker container is converted to a Singularity container on the cluster. With singularity, the container can be executed without root access. We use the cluster's job scheduler, \emph{e.g.}\ slurm, to submit the job within the container.}
    \label{fig:workflow}
\end{figure}

The simplest way to obtain the singularity file would be to prepare a file \lstinline[language=bash]{spack.yaml} and use \lstinline[language=bash]{spack containerize > octotiger.def} to generate the instruction file \lstinline[language=bash]{octotiger.def}. After that, the command \lstinline[language=bash]{singularity build --fakeroot octotiger.sif octotiger.def} builds the singularity image. However, we encountered issues with the generated instruction file on Supercomputer\ Fugaku. One issue was that on Super\ Computer Fugaku CentOS 8 is required to import the RPM packages for the Fujitsu compiler provided by Riken. However, as of this writing, Spack supports only CentOS 7.\footnote{\url{https://spack.readthedocs.io/en/latest/containers.html}}. We used the workflow in Figure~\ref{fig:workflow}. First, we generate a Docker file where we specify the operating system of the Docker image and use our new Spack package to compile Octo-Tiger. We use singularity to convert the Dockerfile into a Singularity file.  We use the job scheduler to run Octo-Tiger within the Singularity image. The same singularity image could be executed in the cloud or the HPC cluster. However, cloud resources were not the scope of this paper. Listing~\ref{lst:instructions} shows the instructions to generate the singularity file from a Docker image and run Octo-Tiger in the image on Supercomputer\ Fugaku.

\begin{lstlisting}[caption=Instructions on Supercomputer\ Fugaku,float=tp,label=lst:instructions:run,language=bash] export TMPDIR=/worktmp
# Generating the image
singularity build -F /worktmp/wamta24.simg docker://stevenrbrandt/wamta24:arm64
# Running the image
singularity exec --bind /worktmp/ /worktmp/wamta24.simg bash octotiger
\end{lstlisting}

\subsection{Challenges in compiling and running within containers}
We identified the following challenge on Supercomputer\ Fugaku. First, on Fugaku the Fusitju compiler and MPI wrapper are available. The user can not easily install these. Riken provides some Docker images and RPM packages to install these packages. However, we use the GCC compiler since the Fusitju compiler does not support C\texttt{++} 17 which is required for Octo-Tiger and HPX. We have the same issue with compiling Octo-Tiger before and this issue is not related to containerization. So adding vendor-specific compilers, \emph{e.g.}\ Cray Compiling Environment (CCE) or IBM's compiler for Power architecture, were not trivial tasks. Second, the Docker image needs to be generated on the host architecture. A challenge was here that we had no access to an A64FX machine with Docker to compile the image. We used Docker's buildx command with support for Linux/arm64 provided by Docker to generate an A64FX-based image. However, Docker's buildx uses cross-compilation for the A64FX-based image. According to the documentation\footnote{\url{https://spack.readthedocs.io/en/latest/features.html}} there is cross-compilation support in Spack. However, we encountered build issues and built Octo-Tiger without Kokkos support using CMake.

\section{Performance differences}
\label{sec:performance:differences}

\begin{table}[tb]
    \centering
    \caption{Statistics for the regular and singularity runs on Supercomputer\ Fugaku. We executed ten runs for each option.}
    \begin{tabular}{l|ccccc}\toprule
         & Min Time  & Median Time  & Average Time  & Max Time  & Standard derivation  \\\midrule
     Singularity &  267.5  & 277.3 & 277.1 & 286.8 & 6.2 \\
     Regular  & 214.1 & 215.2 & 221.1 & 237.6 & 9.8
   \\\bottomrule
    \end{tabular}
    \label{tab:results:fugaku}
\end{table}

There is no free lunch. To investigate the overheads introduced by Singularity images on HPC clusters, we compiled the same version of Octo-Tiger and all its dependencies using spack. The versions used on both supercomputers are documented in the supplementary materials using \lstinline{spack concretize}. First, we compile Octo-Tiger without any container on the node. Second, we compile Octo-Tiger within Docker. After that, we run the rotating star example on the node directly and within the image.

\subsection{Supercomputer\ Fugaku (A64FX)}
On Supercomputer\ Fugaku, we only present single-node scaling. For the distributed runs additional effort was required to add Fusitju MPI with Tofu-D support to the Singularity container. Table~\ref{tab:results:fugaku} shows the statistics out of ten regular and ten singularity runs. We observed that the regular runs were on average 50 seconds faster. The standard derivation is slightly higher for the regular runs.

\subsection{DeepBayou}

\begin{figure}[tb]
    \centering
\subfloat[\label{fig:difference:deepbayou:cpu}]{
\begin{tikzpicture}[scale=0.8]
\begin{axis} [xbar = .05cm,
	bar width = 12pt,
	xmin = 0, 
	xmax = 3146.51, 
	enlarge y limits = {abs = .8},
	enlarge x limits = {value = .25, upper},
    title=CPU,
    grid,
    xlabel=Time (s),
    ylabel= \# Nodes,
    ytick={0,...,5},
    yticklabels={1,2,4,8},
]
\addplot[fill=gray!45] coordinates {(3146.51,0) (907.257,1) (447.791
,2) (220.746
,3) };
\addplot[fill=black] coordinates {(3182.21,0) (808.228,1) (447.156
,2) (139.437,3)  };
\legend{Singularity,Regular};
\end{axis}
\end{tikzpicture}
    }
\subfloat[\label{fig:difference:deepbayou:cuda}]{
\begin{tikzpicture}[scale=0.8]
\begin{axis} [xbar = .05cm,
	bar width = 12pt,
	xmin = 0, 
	xmax = 120, 
	enlarge y limits = {abs = .8},
	enlarge x limits = {value = .25, upper},
    title=CPU + GPU,
    grid,
    xlabel=Time (s),
    ylabel= \# Nodes,
    ytick={0,...,3},
    yticklabels={1,2,4,8},
]
\addplot[fill=gray!45] coordinates {(116.94242857142858,0) (38.441542857142856,1) (19.11595,2) (12.67115,3) };
\addplot[fill=black] coordinates {(108.49182857142857,0) (70.44953333333335,1) (33.66081666666667,2) (17.80804444444444,3) };
\legend{Singularity,Regular};
\end{axis}
\end{tikzpicture}

}
\caption{Comparison of regular and singularity runs on DeepBayou using CPUs \protect\subref{fig:difference:deepbayou:cpu} and CPUs + GPUs \protect\subref{fig:difference:deepbayou:cuda}.}
\label{fig:enter-label}
\end{figure}
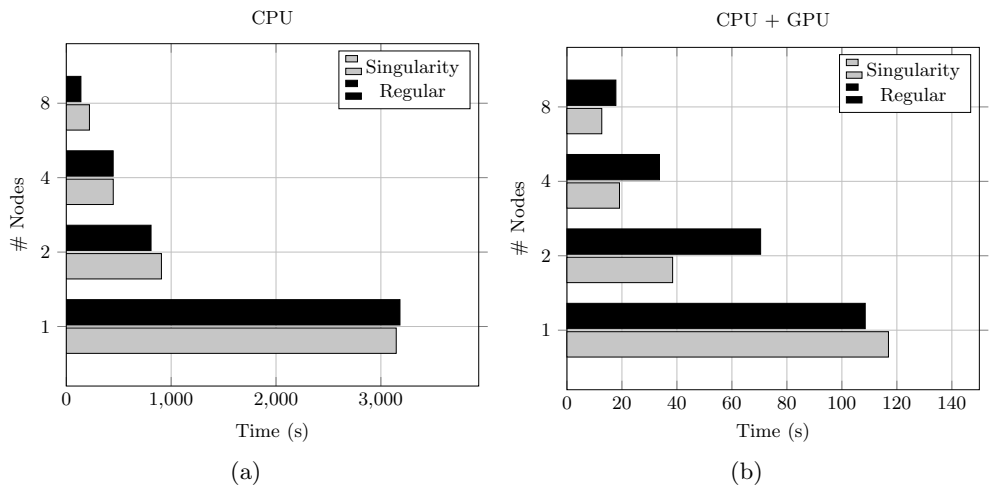

Deep Bayou is a small GPU cluster hosted at LSU with two 24-core Intel Xeon Gold 6248R CPUs and two NVIDIA V100S GPUs per node. Here, we used spack to compile Octo-Tiger with Kokkos support for NVIDIA GPUs. Figure~\ref{fig:difference:deepbayou:cpu} shows the measurements using the CPUs without GPUs. For most runs, we observe that the computation time is comparable. Figure~\ref{fig:difference:deepbayou:cuda} shows the measurements on a single node up to eight nodes using the CPUs and GPUs. For the CPU runs both versions are on par, however, for the GPU runs the singularity runs were faster. This has been observed for other runs as well~\cite{8950978}. We are using the default performance parameters without further tuning them for the platform or GPU architecture. We note that it was far less challenging to get the code running on Deep Bayou as we were able to use Spack for both the image and the native build to compile.

The CPU-only runs with one node appear to be an anomaly: Here we are slower than expected, both for the regular run and for the singularity run.
However, this does only happen in the CPU runs with one node, not in any of the other runs. The other CPU runs with two or more nodes running as expected (causing a super linear speedup when going from one node to two nodes). We are currently investigating this issue, especially because we do not see this on any of the other platforms we currently use Octo-Tiger.

\section{Conclusion and Outlook}
\label{sec:conclusion}
This paper set out to evaluate the overheads of containers when using HPX/Kokkos. However, to get to this point was quite a journey. One important lesson learned was that the workflow in Figure~\ref{fig:workflow} is not always viable.
In the Docker container, vendor-specific compilers, MPI wrappers, or libraries are not installed. For nonconventional architecture, like A64FX, it can be a challenge to build a Singularity or Docker image.
We were unable to successfully build a Singularity image directly on Fugaku, however, we were able to build a docker image using buildx (but could not get Spack to work in that environment). On Queen Bee 3 (qbc) we were able to build with Spack both inside and outside the container, but found we needed to use MPICH instead of OpenMPI to properly interface with LSU Slurm.
After struggling with building, on the other hand, running the containers was a walk in the park. For the performance difference due to potential overheads, we observed that on Supercomputer\ Fugaku the regular runs were faster. On DeepBayou for CPU runs, there was not much difference. However, for CPU and GPU runs, we identified some differences on a single node, but all distributed runs in the container crashed.

We conclude that containers offer benefits for reproducibility, porting, and running HPC applications, but building in a container is not always easier. While performance on Fugaku did benefit from running outside the container, the gains in reproducibility and documentation of the build process probably outweigh them.

For future work, we would to get the Fusitju MPI to interface with the container and do distributed runs.
For the runs with GPUs, more investigation into why we observe the performance difference is needed. In addition, we would expand our study to larger GPU-based supercomputers such as NERSC's Perlmutter.

\section*{Acknowledgments}
{\footnotesize
Funded partly by NSF \#229751: POSE: Phase 1: Constellation: A Pathway to Establish the STE||AR Open-Source Organization. 
Computational resources of the Supercomputer\ Fugaku provided by the RIKEN Center for Computational Science were used.
}

\section*{Supplementary materials}
{\footnotesize
All scripts are available on GitHub (\url{https://github.com/diehlpkpapers/dockerizeHPX}).
}

%
%
%
{\footnotesize
\bibliographystyle{splncs04}
\bibliography{software}

\begin{thebibliography}{10}
\providecommand{\url}[1]{\texttt{#1}}
\providecommand{\urlprefix}{URL }
\providecommand{\doi}[1]{https://doi.org/#1}

\bibitem{9284294}
Abraham, S., et~al.: {On the Use of Containers in High Performance Computing Environments}. In: 2020 IEEE 13th International Conference on Cloud Computing. pp. 284--293 (2020)

\bibitem{alles2018assessing}
Alles, G.R., et~al.: {Assessing the computation and communication overhead of Linux containers for HPC applications}. In: 2018 Symposium on High Performance Computing Systems. pp. 116--123. IEEE (2018)

\bibitem{7923813}
Azab, A.: {Enabling Docker Containers for High-Performance and Many-Task Computing}. In: 2017 IEEE International Conference on Cloud Engineering. pp. 279--285 (2017)

\bibitem{bauer2012legion}
Bauer, M., et~al.: Legion: Expressing locality and independence with logical regions. In: SC'12: Proceedings of the International Conference on High Performance Computing, Networking, Storage and Analysis. pp. 1--11. IEEE (2012)

\bibitem{7923810}
de~Bayser, M., et~al.: {Integrating MPI with Docker for HPC}. In: 2017 IEEE International Conference on Cloud Engineering. pp. 259--265 (2017)

\bibitem{benedicic2019sarus}
Benedicic, L., et~al.: {Sarus: Highly scalable docker containers for HPC systems}. In: High Performance Computing: ISC High Performance 2019 International Workshops, Frankfurt, Germany, June 16-20, 2019, Revised Selected Papers 34. pp. 46--60. Springer (2019)

\bibitem{bosilca2013parsec}
Bosilca, G., et~al.: Parsec: Exploiting heterogeneity to enhance scalability. Computing in Science \& Engineering  \textbf{15}(6),  36--45 (2013)

\bibitem{casalicchio2017measuring}
Casalicchio, E., Perciballi, V.: {Measuring docker performance: What a mess!!!} In: Proceedings of the 8th ACM/SPEC on International Conference on Performance Engineering Companion. pp. 11--16 (2017)

\bibitem{chamberlain2007parallel}
Chamberlain, B.L., et~al.: Parallel programmability and the chapel language. The International Journal of High Performance Computing Applications  \textbf{21}(3),  291--312 (2007)

\bibitem{7562612}
Chung, M.T., Quang-Hung, et~al.: {Using Docker in high performance computing applications}. In: 2016 IEEE Sixth International Conference on Communications and Electronics. pp. 52--57 (2016)

\bibitem{chung2016using}
Chung, M.T., et~al.: Using docker in high performance computing applications. In: 2016 IEEE Sixth International Conference on Communications and Electronics. pp. 52--57. IEEE (2016)

\bibitem{9882991}
Courtes, L.: {Reproducibility and Performance: Why Choose?} Computing in Science \& Engineering  \textbf{24}(03),  77--80 (2022)

\bibitem{10.1145/3295500.3356221}
Dai\ss{}, G., et~al.: {From Piz Daint to the Stars: Simulation of Stellar Mergers Using High-Level Abstractions}. In: {Proceedings of the International Conference for High Performance Computing, Networking, Storage and Analysis}. SC '19, ACM, New York, NY, USA (2019)

\bibitem{daiss2021beyond}
Dai{\ss}, G., et~al.: {Beyond fork-join: Integration of performance portable Kokkos kernels with HPX}. In: 2021 IEEE International Parallel and Distributed Processing Symposium Workshops. pp. 377--386. IEEE (2021)

\bibitem{10.1145/3585341.3585354}
Dai\ss{}, G., et~al.: Stellar mergers with hpx-kokkos and sycl: Methods of using an asynchronous many-task runtime system with sycl. In: Proceedings of the 2023 International Workshop on OpenCL. ACM, New York, NY, USA (2023)

\bibitem{daiss2022simd}
Daiß, G., et~al.: {From Merging Frameworks to Merging Stars: Experiences using HPX, Kokkos and SIMD Types}. In: 2022 IEEE/ACM 7th International Workshop on Extreme Scale Programming Models and Middleware. pp. 10--19. IEEE, Los Alamitos, CA, USA (2022)

\bibitem{daiss2022aggregation}
Daiß, G., et~al.: {From Task-Based GPU Work Aggregation to Stellar Mergers: Turning Fine-Grained CPU Tasks into Portable GPU Kernels}. In: 2022 IEEE/ACM International Workshop on Performance, Portability and Productivity in HPC. pp. 89--99. IEEE, Los Alamitos, CA, USA (2022)

\bibitem{diehl2023simulating}
Diehl, P., et~al.: {Simulating Stellar Merger using HPX/Kokkos on A64FX on Supercomputer Fugaku} (2023)

\bibitem{gamblin2015spack}
Gamblin, T., et~al.: {The Spack package manager: bringing order to HPC software chaos}. In: Proceedings of the International Conference for High Performance Computing, Networking, Storage and Analysis. pp. 1--12 (2015)

\bibitem{germain2000uintah}
Germain, J.D.d.S., et~al.: Uintah: A massively parallel problem solving environment. In: Proceedings the Ninth International Symposium on High-Performance Distributed Computing. pp. 33--41. IEEE (2000)

\bibitem{kaiser2020hpx}
Hartmut, K., et~al.: {HPX-the C\texttt{++} standard library for parallelism and concurrency}. Journal of Open Source Software  \textbf{5}(53), ~2352 (2020)

\bibitem{10.1007/978-3-319-20119-1_36}
Higgins, J., et~al.: {Orchestrating Docker Containers in the HPC Environment}. In: Kunkel, J.M., Ludwig, T. (eds.) High Performance Computing. pp. 506--513. Springer, Cham (2015)

\bibitem{kadam2018numerical}
Kadam, K., et~al.: Numerical simulations of mass transfer in binaries with bipolytropic components. MNRAS  \textbf{481}(3),  3683--3707 (2018)

\bibitem{kale1993charm++}
Kale, L.V., Krishnan, S.: {Charm\texttt{++} a portable concurrent object oriented system based on C\texttt{++}}. In: Proceedings of the eighth annual conference on Object-oriented programming systems, languages, and applications. pp. 91--108 (1993)

\bibitem{7921010}
Li, Z., et~al.: {Performance Overhead Comparison between Hypervisor and Container Based Virtualization}. In: 2017 IEEE 31st International Conference on Advanced Information Networking and Applications. pp. 955--962 (2017)

\bibitem{marcello2021octo}
Marcello, D.C., et~al.: {Octo-Tiger: a new, 3D hydrodynamic code for stellar mergers that uses HPX parallelization}. {MNRAS}  \textbf{504}(4),  5345--5382 (2021)

\bibitem{merkel2014docker}
Merkel, D., et~al.: Docker: lightweight linux containers for consistent development and deployment. Linux j  \textbf{239}(2), ~2 (2014)

\bibitem{9547695}
Plale, B.A., Malik, T., Pouchard, L.C.: Reproducibility practice in high-performance computing: Community survey results. Computing in Science \& Engineering  \textbf{23}(05),  55--60 (sep 2021)

\bibitem{rad2017introduction}
Rad, B.B., et~al.: An introduction to docker and analysis of its performance. International Journal of Computer Science and Network Security  \textbf{17}(3), ~228 (2017)

\bibitem{8748885}
Rezende~Alles, G., et~al.: {Assessing the Computation and Communication Overhead of Linux Containers for HPC Applications}. In: 2018 Symposium on High Performance Computing Systems. pp. 116--123 (2018)

\bibitem{8820966}
Rudyy, O., et~al.: {Containers in HPC: A Scalability and Portability Study in Production Biological Simulations}. In: 2019 IEEE International Parallel and Distributed Processing Symposium. pp. 567--577 (2019)

\bibitem{10.1145/3219104.3229280}
Saha, P., et~al.: {Evaluation of Docker Containers for Scientific Workloads in the Cloud}. In: Proceedings of the Practice and Experience on Advanced Research Computing. PEARC '18, ACM, New York, NY, USA (2018)

\bibitem{sahasrabudhe2019portable}
Sahasrabudhe, D., et~al.: {A portable SIMD primitive using Kokkos for heterogeneous architectures}. In: {International Workshop on Accelerator Programming Using Directives}. pp. 140--163. Springer (2019)

\bibitem{sparks2019enabling}
Sparks, J.: {Enabling docker for HPC}. Concurrency and Computation: Practice and Experience  \textbf{31}(16),  e5018 (2019)

\bibitem{staff2018role}
Staff, J.E., et~al.: The role of dredge-up in double white dwarf mergers. The Astrophysical Journal  \textbf{862}(1), ~74 (2018)

\bibitem{thoman2018taxonomy}
Thoman, P., et~al.: A taxonomy of task-based parallel programming technologies for high-performance computing. The Journal of Supercomputing  \textbf{74}(4),  1422--1434 (2018)

\bibitem{9485033}
Trott, C.R., et~al.: {Kokkos 3: Programming Model Extensions for the Exascale Era}. IEEE Transactions on Parallel and Distributed Systems  \textbf{33}(4),  805--817 (2022)

\end{thebibliography}
}

\end{document}